# Quantum optimization algorithms for CT image segmentation from X-ray data

Kyungtaek Jun*[,a]
[a]Research Center, Qtomo, Gyeonggi-do, 14548, South Korea

[*]Corresponding author: ktfriends@gmail.com

## ABSTRACT

Computed tomography (CT) is an important imaging technique used in medical analysis of the internal structure of the human body. Previously, image segmentation methods were required after acquiring reconstructed CT images to obtain segmented CT images which made it susceptible to errors from both reconstruction and segmentation algorithms. However, this paper introduces a new approach using an advanced quantum optimization algorithm called quadratic unconstrained binary optimization (QUBO) for CT image segmentation. This algorithm allows CT image reconstruction and segmentation to be performed simultaneously. This algorithm segments CT images by minimizing the difference between a sinogram in a superposition state with qubits, obtained using the mathematical projection including the Radon transform, and the experimentally acquired sinogram from X-ray images for various angles. Furthermore, we leveraged X-ray mass attenuation coefficients to reduce the number of logical qubits required for our quantum optimization algorithm, and we employed D-Wave's hybrid solver to solve the optimization problem. We compared the segmentation results of our algorithm with those of classical algorithms using X-ray images of actual tooth samples to validate the results of our algorithm. The comparison revealed that, after undergoing appropriate image post-processing, our algorithm's segmentation results matched those of classical algorithms that perform segmentation after reconstruction, except for some pixels at the boundary. Moreover, since our algorithm focuses on each pixel in the sinogram, it effectively generates QUBOs in regions where significant noise has been removed, leading to successful outcomes. Thanks to this feature of our algorithm, it delivers strong results even with data obtained from limited angles. Therefore, to demonstrate quantum supremacy, we conducted experiments under conditions of significant noise and limited-angle data separately. These experiments yielded successful and meaningful results. We expect that the new quantum optimization tomographic reconstruction algorithm will bring about great advancements in medical imaging and electron tomography.

## 1. INTRODUCTION

Quantum supremacy refers to the milestone in quantum computing where a quantum computer performs a calculation that is infeasible for classical computers to execute in a reasonable time frame. This concept, first formally introduced by physicist John Preskill in 2012, marks a pivotal point in the field of quantum computing, showcasing the potential for quantum processors to solve certain types of problems significantly faster than their classical counterparts[1]. As research has progressed, the practical demonstration of quantum supremacy has become a central goal, leading to the exploration of various algorithms and computational tasks that could highlight the advantages of quantum over classical systems. The examples to demonstrate quantum supremacy include Integer factorization[2], the boson sampling[3], quantum random circuit sampling[4,5], Quantum Approximate Optimization Algorithm (QAOA)[6], etc. These algorithms not only serve as benchmarks for quantum advantage but also drive the development of quantum optimization techniques, which are becoming increasingly critical as quantum hardware advances. The advancements in quantum computing have made quantum optimization algorithms increasingly important. Recent developments in various scientific fields[7,8,9] have led to the creation of quadratic unconstrained binary optimization (QUBO) models specifically designed for these algorithms. The QUBO model is well-suited for utilization in gate model-based quantum computers, particularly when combined with the QAOA[10,11]. To maximize performance when working with QUBO models, it is advisable to employ a quantum annealer instead of a gate model quantum computer, as it offers superior performance compared to a gate model quantum computer. D-Wave's Advantage quantum annealer currently stands out with its impressive specifications, boasting over 5,000 qubits and more than 35,000 couplers. This allows for the use of 180 logical qubits with interconnections. Notably, D-Wave also offers a hybrid model that supports up to one million variables and 200 million biases.

Computed tomography (CT) is a powerful tool for non-destructive analysis of an object's internal structure, enabling detailed examination without causing any damage to the object. To obtain a segmented image from X-ray projection data, the following two steps are required. First, the CT image is reconstructed by applying the back-projection algorithm to the X-ray projection data. By applying a segmentation algorithm to the reconstructed CT image, we can obtain a segmentation image. CT employs a back-projection algorithm to reconstruct an object's internal structure based on projected images captured from different angles. An inverse projection utilizes various algorithms; iterative methods[12], fast Fourier transform (FFT)[13], artificial intelligence[14], and optimization[15] to name a few examples. In particular, the optimization algorithm generates more accurate images similar to the actual internal structure by utilizing the complete sinogram pattern rather than using the sinogram rows sequentially[16]. By iteratively updating the sinogram using the Radon transform on a randomly generated CT image, this algorithm minimizes the disparity between the reconstructed CT image and the true sinogram. However, the classical algorithm associated with classical computers often hinders the quality of the obtained CT images. Moreover, data errors in X-ray images can arise due to noise generated during object scanning, leading to various artifacts such as ring artifacts and motion artifacts. To address this challenge, the QUBO model can be implemented on a quantum computer, leveraging its computational efficiency and resemblance to the optimization algorithm[10]. Nevertheless, improving CT image quality remains a hurdle within the QUBO model due to the limited number of logical qubits and the low probability of discovering the minimum energy state in a quantum computer. In the case of calculating the QUBO model for CT image reconstruction using D-Wave's hybrid solver, a solution or a good approximation was obtained even when using about 10,000 logical qubits[17].

In this paper, we introduce the QUBO model for quantum segmentation that can perform both CT image reconstruction and segmentation. We assume that the X-ray mass attenuation of the sample is already known. The new algorithm begins by representing the X-ray mass attenuations as a quantum superposition of the qubits for each pixel of the CT image. Then each pixel becomes a quantum superposition representing all the values in the sample's materials and space. Afterward, the new algorithm obtains superposed projection data by applying the same mathematical algorithm as the X-ray beam shape obtained from the X-ray projection image to the CT image in the superposed state. For each pixel, the difference between the superposed projection image and the X-ray projection image is calculated in the form of QUBO as a least-squared problem. Finally, the QUBO model can be formulated by summing the QUBO form for each pixel of all projection images. We verify the new algorithm for a parallel beam X-ray light source. The sizes of the images used in the experiments were $50 \times 50$ and $100 \times 100$, with a total of 2,500 and 10,000 logical qubits, respectively. The mathematical projection used for superposed CT images is the Radon transform. The global minimum energy of the QUBO model was obtained using the hybrid solver of D-Wave system. We verified using a parallel beam type X-ray light source in this paper, but it is possible to use the new algorithm for any type of X-ray light source. We anticipate that medical imaging diagnostic technology will advance significantly when the new quantum optimization tomographic reconstruction algorithm is applied to medical CT systems. Furthermore, to demonstrate the quantum supremacy of our algorithm, we conducted several experiments. We compared the results of classical and quantum reconstruction algorithms on CT images, focusing particularly on cases where the projection data contained errors or noise. We included results from our new algorithm applied to projection data that created ring artifacts in the reconstructed images. Additionally, to assess whether our algorithm could optimize the missing wedge of information in electron tomography, we performed experiments using projection data with restricted angles[18]. These results underscore the potential of our approach not only in medical imaging but also in its broader applicability across various fields of tomography.

# 2. BACKGROUND

## 2.1 Radon transform and sinogram

The Radon transform mathematically describes how projection data for a parallel-beam type X-ray light source is obtained. The Radon transform can be calculated as shown:

$$R(p,\tau)[(x,y)] = \int_{-\infty}^{\infty} f(x,\ \tau + px)dx \quad (1)$$

$$= \int_{-\infty}^{\infty}\int_{-\infty}^{\infty} f(x,y)\,\delta[y - (\tau + px)]dydx \quad (2)$$

where $p$ is the slope of a line, $\tau$ is its intercept, and $\delta(x)$ is the delta function. A sinogram is created by accumulating Radon transforms according to angles. When using the same sample with different positions, the projected position by the Radon transform will be different. In a sinogram without motion artifacts, the shape of the projected object is said to be an ideal sinogram pattern, which is an important factor when reconstructing a CT image.

## 2.2 Classical optimization algorithm

Since 2017, researchers have developed an alignment method for correcting motion artifacts in the entire projection image during scanning[19,20,21,22]. This method, unlike previous approaches, satisfies the Helgason-Ludwig consistency condition by calculating points within the sample in a virtual space. Consequently, it enables the acquisition of high-quality CT images even when the center of rotation in space is changed. In 2021, an algorithm was introduced for reconstructing CT images based on sinogram patterns. Notably, this algorithm allows for controlling the reconstructed position of the sample within the CT image by manipulating the sinogram pattern[9]. By randomly reconstructing the CT images within a specific region of interest, the algorithm minimizes the discrepancy between the sinogram generated from the initial CT image using the Radon transform and the actual sinogram. Moreover, as it employs the complete sinogram pattern of the sample, it exhibits robustness against artifacts from specific angles. The CT image reconstruction algorithm can be mathematically formulated as an optimization problem, outlined as follows:

$$\underset{T}{\operatorname{argmin}}\ \mathrm{MSE}(R(T), S) \quad (3)$$

where $T$ is one of the CT images, $S$ is the given sinogram, and $R$ is the Radon transform. PyTorch (1.9.0+cu111) was used to calculate the Radon transform, and the Adam optimizer of the PyTorch library was used to calculate the optimization.

## 2.3 QUBO and Ising models

QUBO is a combinatorial optimization problem and its global minimization energy can be calculated in a quantum annealer. The set of binary vectors of a fixed length $n > 0$ is denoted by $\mathbb{B}^n$, where $\mathbb{B} = \{0,1\}$ is the set of binary values. A cost function $f$ defined within an n-dimensional binary vector space, $\mathbb{B}^n$, on real numbers, $\mathbb{R}$.

$$f(\vec{q}) = \sum_{i=1}^{n} Q_{i,i}q_i + \sum_{i<j}^{n} Q_{i,j}q_iq_j \quad (4)$$

where $\vec{q}$ is an element of $\mathbb{B}^n$. If there is an upper triangular matrix $Q \in \mathbb{R}^{n\times n}$ with $Q_{i,i}$ as the $(i,i)$ component and $Q_{i,j}$ as the $(i,j)$ component, it can be simply expressed as follows:

$$f(\vec{q}) = \vec{q}^T Q\vec{q} \quad (5)$$

Within matrix $Q$, the diagonal elements $Q_{i,i}$ symbolize the linear terms, whereas off-diagonal elements $Q_{i,j}$ represent the quadratic terms. The QUBO problem consists of finding a binary vector $\vec{q}^*$that is minimal with respect of $f$ as follows:

$$f(\vec{q}^*) \leq \vec{q}^T Q \vec{q} \tag{6}$$

The Ising model has $\sigma_i \in \{-1,1\}$ as an element instead of $q_i$. The Ising model can be converted from the QUBO model through linear transformation.

$$\sigma \rightarrow 2q - 1 \text{ or } q \rightarrow \frac{1}{2}(\sigma + 1) \tag{7}$$

### 2.4 QUBO formulation for linear systems

For a matrix $A \in R^{n\times n}$, given a column vector of variables $\vec{x} \in R^n$ and a column vector $\vec{b} \in R^n$, the objective is to determine $\vec{x}$ such that $A\vec{x} = \vec{b}$. The objective of the linear least-squares problem is to identify the vector $\vec{x}^*$ that minimizes the norm $\left\|A\vec{x} - \vec{b}\right\|$, which can be expressed as follows:

$$\arg\min_{\vec{x}} \left\|A\vec{x} - \vec{b}\right\| \tag{8}$$

The non-negative integer value $x_i$ can be denoted by the radix-2 representation as follows:

$$x_i \approx \sum_{l=0}^{m-1} 2^l \, q_{i,l} \tag{9}$$

To formulate Eq. (8) into the QUBO model, we start the calculation from $\| A\vec{x} - \vec{b} \|_2^2$.

$$\| A\vec{x} - \vec{b} \|_2^2 = \sum_{k=1}^{n} \left\{ \sum_{i=1}^{n} \left(a_{k,i} x_i\right)^2 + 2 \sum_{i<j} a_{k,i} \, a_{k,j} x_i x_j - 2b_k \sum_{i=1}^{n} a_{k,i} \, x_i + {b_k}^2 \right\} \tag{10}$$

Each term of Eq. (10) is calculated as follows:

$$\sum_{k=1}^{n}\sum_{i=1}^{n} \left(a_{k,i} x_i\right)^2 \approx \sum_{k=1}^{n}\sum_{i=1}^{n}\sum_{l=0}^{m-1} a_{k,i}^2 \, 2^{2l} q_{i,l} + \sum_{k=1}^{n}\sum_{i=1}^{n}\sum_{l_1<l_2} a_{k,i}^2 \, 2^{l_1+l_2+1} q_{i,l_1} q_{i,l_2} \tag{11}$$

$$\sum_{k=1}^{n}\sum_{i<j} 2 \, a_{k,i} a_{k,j} x_i x_j \approx \sum_{k=1}^{n}\sum_{i<j}\sum_{l_1=0}^{m-1}\sum_{l_2=0}^{m-1} 2^{l_1+l_2+1} \, a_{k,i} a_{k,j} q_{i,l_1} q_{j,l_2} \tag{12}$$

$$\sum_{k=1}^{n}\sum_{i=1}^{n} \left(-2a_{k,i} b_k x_i\right) \approx \sum_{k=1}^{n}\sum_{i=1}^{n}\sum_{l=0}^{m-1} 2^{l+1} \, a_{k,i} b_k \left(-q_{i,l}\right) \tag{13}$$

The QUBO model for a linear system with a positive solution is the sum of the three aforementioned terms. The global minimum energy of the QUBO model is $-\vec{b}^T \cdot \vec{b}$. The pseudocode for the QUBO model is shown in Algorithm 1.

**Algorithm 1** Calculating the QUBO matrix for QUBO model

Data input $A, \vec{b}$, and qubit numbers $m$ for each element of $\vec{x}$

**for** $k = 1{:}n$ **do**

    **for** $i = 1{:}n$ **do**
        **for** $l = 0{:}m{-}1$ **do**
            $Q(m(i-1)+l+1, m(i-1)+l+1) \leftarrow 2^{2l}(A(k,i))^2 - 2^{l+1}A(k,i)b(k)$

**for** $k = 1{:}n$ **do**

    **for** $i = 1{:}n$ **do**
        **for** $l_1 = 0{:}m-2$ **do**

            **for** $l_2 = l_1+1{:}m-1$ **do**
                $Q(m(i-1)+l_1+1, m(i-1)+l_2+1) \leftarrow 2^{l_1+l_2+1}(A(k,i))^2$

**for** $k = 1{:}n$ **do**

    **for** $i = 1{:}n-1$ **do**

        **for** $j = i+1{:}n$ **do**
            **for** $l_1 = 0{:}m-1$ **do**
                **for** $l_2 = 0{:}m-1$ **do**
                    $Q(m(i-1)+l_1+1, m(j-1)+l_2+1) \leftarrow 2^{l_1+l_2+1}A(k,i)A(k,j)$

---

### 2.5 QUBO formulation for CT image reconstruction

Quantum annealers and gate model quantum computers offer the QUBO and Ising models as quantum optimization algorithms. In quantum annealing, the model's minimum value is computed, while in gate model quantum computers, the maximum value is calculated using QAOA. This paper focuses on discussing the utilization of QUBO modeling for minimization.

Typically, the pixel value in the projection image correlates with the X-ray intensity passing through the sample's thickness[10]. In regions where the Beer-Lambert law applies, we can express the value at the i-th position, denoted as $P_i$ for a specific axial level, using Eq. 14.

$$P_i = \int_i^{i+1} S(x,y)dl \tag{14}$$

where $S(x,y)$ is the X-ray mass attenuation coefficient for the $(x,y)$-position of the sample, and the $l$ axis is defined perpendicular to the direction of the X-ray. This equation can be expressed as a sum of discrete forms as shown in Eq. 15.

$$P_i \approx \sum a_k T_k \tag{15}$$

where $a_k$ and $T_k$ are the positions and proportions of pixels in the CT image that affect $P_i$, respectively. We may assume that $T_k$ is the average X-ray mass attenuation coefficient for the position $T_{ij}$ of the CT image which affects $P_i$. $S(\theta, i)$ is equal to $P_i$ for the projection angle $\theta$, as a sinogram $S$ is an accumulation of $P$ along the projection angle. We do not know the value of $T_{ij}$. In order to formulate the QUBO model, $T_{ij}$ is expressed as the summation of qubit variables $\sum_{k=0}^{m} 2^k q_k^{ij}$ . The QUBO model for each pixel in the sinogram is computed as follows:

$$\{(P-S)(\theta,i)\}^2 = \left\{\sum_k a_k T_k - S(\theta,i)\right\}^2 \tag{16}$$

$$= \left(\sum_k a_k T_k\right)^2 - 2S(\theta,i)\left(\sum_k a_k T_k\right) + \{S(\theta,i)\}^2 \tag{17}$$

$$= \sum_k {a_k}^2 T_k + 2\sum_{k_1<k_2} a_{k_1} a_{k_2} T_{k_1} T_{k_2} - 2S(\theta,i)\left(\sum_k a_k T_k\right) + \{S(\theta,i)\}^2 \tag{18}$$

Therefore, the QUBO model for CT image reconstruction is calculated as below:

$$\sum_{\theta=0}^{180-d\theta} \sum_{i=1}^{n} \{(P-S)(\theta,i)\}^2 - \sum_{\theta=0}^{180-d\theta} \sum_{i=1}^{n} \{S(\theta,i)\}^2 \tag{19}$$

### 2.6 Advantages of the quantum optimization tomographic reconstruction algorithm

The quantum optimized CT algorithm uses a quantum linear system to formulate the QUBO model. The existing algorithm requires $O(n^3)$ to obtain the solution of an n by n linear system, but $O(n^2)$ is required to formulate the QUBO model for a quantum linear system. In addition, the QUBO formulation can efficiently reduce the amount of calculation to $O(n)$ and up to $O(\log_2 n)$ through parallel computation. Being able to quickly calculate speed through quantum algorithms is also a huge advantage. A greater advantage of the quantum optimized CT algorithm is that clean CT image reconstruction is possible through global minimum energy. The left panel in Fig. 1 shows the classical algorithm starting from the blue dot and approaching the local minimum energy, which is the red dot, through an iterative algorithm. The right panel in Fig. 1 is an illustration using the quantum optimization algorithm. Through the quantum superposition phenomenon, the energies are expressed as blue and red dots, and then the red dot is finally selected through global energy minimum.

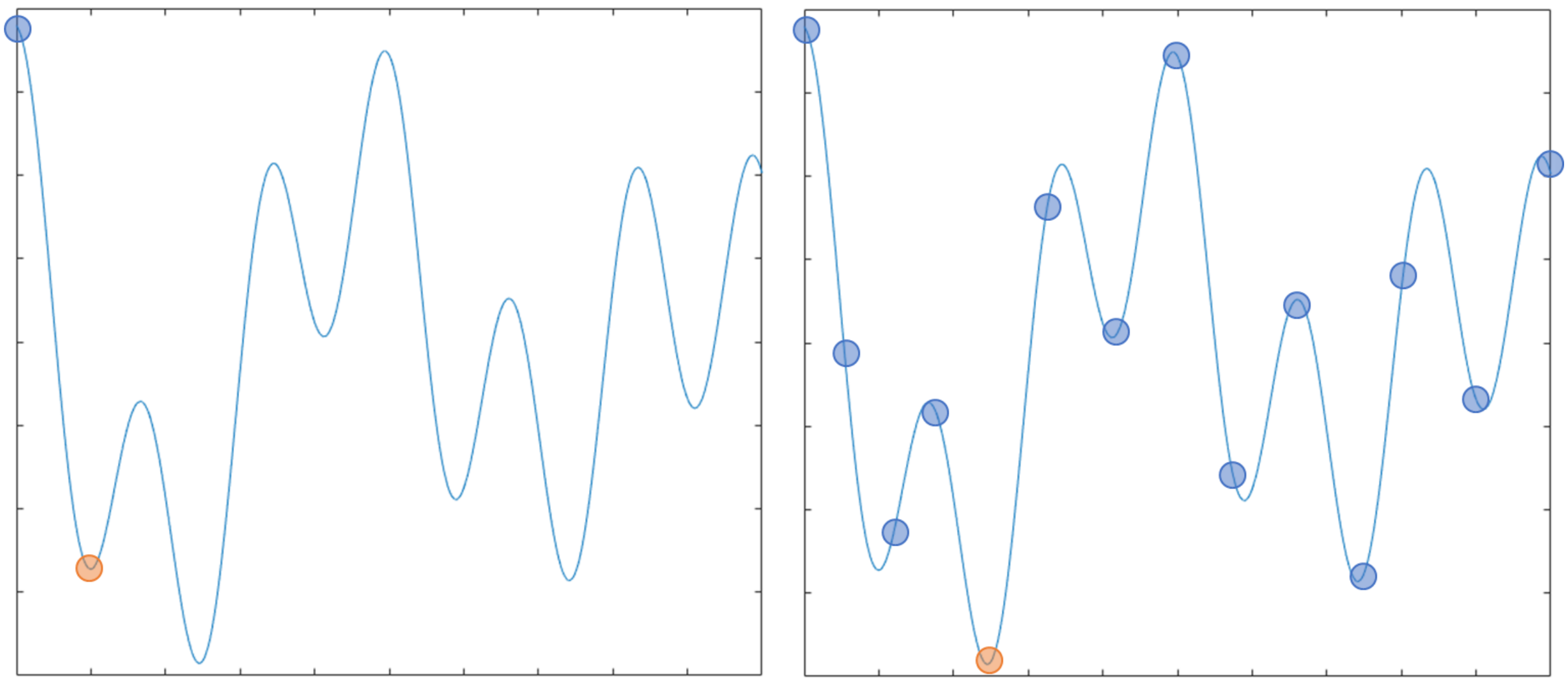

Figure 1. Energy optimization graph. The x-axis represents the values of the n by n array that makes up the CT image, and the y-axis represents the energy difference at that time. The left panel shows the classical algorithm starting from a blue point and energy minimizing to a red point. The right panel shows that the quantum optimization tomographic reconstruction algorithm superposed the energies of the blue and red dots and ultimately obtains the global minimum energy to the red dot.

## 3. METHOD

### 3.1 Data acquisition

An artificial molar that has been processed to attach crowns are made of amalgam (see in Fig. 2a). X-ray microcomputed tomography (μCT) was performed at beamline 6C BioMedical Imaging at the Pohang Light Source-II[23]. The projection image in Fig. 2b at $2560 \times 2160$ pixels resolution was obtained and converted to $50 \times 45$ pixels resolution in Fig. 2c for experimental purposes. We did $50 \times 50$ binning around the center of rotation. Each projection image set was measured every 0.5° degrees, a total of 360 projection images per set.

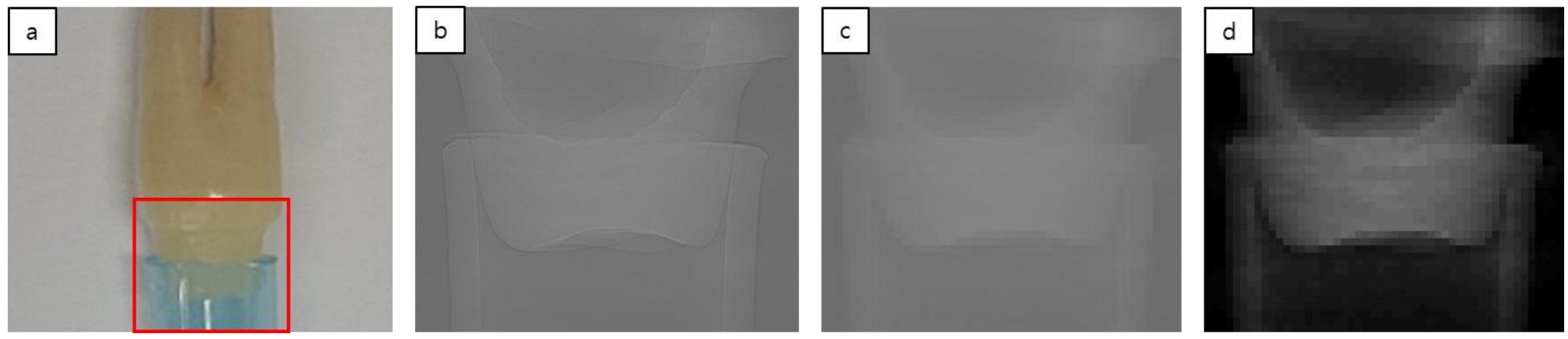

Figure 2. Acquisition and conversion of X-ray projection image. (a) An artificial molar made of amalgam and CT scan inside the red border. (b) Original X-ray projection image with $2560 \times 2160$ pixels resolution. (c) Binned X-ray projection image with $50 \times 45$ pixels resolution. (d) X-ray projection image with the projected area of space set to 0.

### 3.2 Quantum segmentation algorithm for the X-ray data

Assume that the sample exists in 3-dimensional space. Let $\alpha = \frac{\mu}{\rho}$ is the X-ray mass attenuation coefficient[24] and a natural number. If the number of different X-ray mass attenuation coefficients each grid space can have been $m$, then each pixel $I_{ij}$ in the reconstructed image is represented by one of combinations of qubits and binary numbers in Eq. 10.

$$I_{ij} = \sum_{k=1}^{m} \alpha_k q_k^{ij} \tag{20}$$

Here, $q_k^{ij}$ is 0 or 1, and $I_{ij}$ have any elements of set $\{0, \alpha_1, \alpha_2, \cdots, \alpha_m\}$ represents the $(i, j)$ position in a 2-dimensional space. In three-dimensional space, use $(i, j, z)$ instead of $(i, j)$ where $z$ is an axial level.

To apply optimization algorithm onto the X-ray projection data in parallel beam type, we use a Radon transform on the superposed CT image. Let $IP$ be the superposed sinogram transformed by the CT image $I$. For the projection angle $\theta$, the s-th position of $IP$ is calculated as in Eq. 21.

$$IP(\theta, s) = \sum_{i,j} c_{ij} I'_{ij} \tag{21}$$

where $I'_{ij}$ denotes the pixel that affects $IP(\theta, s)$ when the CT image is projected and $c_{ij}$ is the overlapping area when $I'_{ij}$ is projected. Applying the least square equation of the difference between $P(\theta, s)$ and $IP(\theta, s)$, the QUBO model is calculated as follows:

$$\left(IP(\theta, s) - P(\theta, s)\right)^2 = \left(\sum_{i,j} c_{ij} I'_{ij} - P(\theta, s)\right)^2 \tag{22}$$

$$= \left( \sum_{i,j} c_{ij} \sum_{k=1}^{m} \alpha_k q_k^{ij} - P(\theta, s) \right)^2 \tag{23}$$

$$= \left( \sum_{i,j} c_{ij} \sum_{k=1}^{m} \alpha_k q_k^{ij} \right)^2 - 2P(\theta, s) \sum_{i,j} c_{ij} \sum_{k=1}^{m} \alpha_k q_k^{ij} + \{P(\theta, s)\}^2 \tag{24}$$

In Eq. 24, the first term is linear and quadratic terms in the QUBO model, and the third term represents a part of the optimization value. The first term is calculated as follows:

$$\left( \sum_{i,j} c_{ij} \sum_{k=1}^{m} \alpha_k q_k^{ij} \right)^2 = \sum_{i,j} \left( c_{ij} \sum_{k=1}^{m} \alpha_k q_k^{ij} \right)^2 + 2 \sum_{\substack{i,i',j,j' \\ i \neq i' or j \neq j'}} \left( c_{ij} \sum_{k=1}^{m} \alpha_k q_k^{ij} \right) \left( c_{i'j'} \sum_{k'=1}^{m} \alpha_{k'} q_{k'}^{i'j'} \right) \tag{25}$$

$$= \sum_{i,j} (c_{ij})^2 \left( \sum_{k=1}^{m} \alpha_k q_k^{ij} \right)^2 + \sum_{\substack{i,i',j,j' \\ i \neq i' or j \neq j'}} \sum_{1 \leq k,k' \leq m} 2 c_{ij} c_{i'j'} \alpha_k \alpha_{k'} q_k^{ij} q_{k'}^{i'j'} \tag{26}$$

By simplifying the first term in Eq. 26, we calculate the following equation.

$$\sum_{i,j} (c_{ij})^2 \left( \sum_{k=1}^{m} \alpha_k q_k^{ij} \right)^2 = \sum_{i,j} (c_{ij})^2 \left\{ \sum_{k=1}^{m} \left( \alpha_k q_k^{ij} \right)^2 + \sum_{1 \leq k < k' \leq m} 2 \alpha_k \alpha_{k'} q_k^{ij} q_{k'}^{ij} \right\} \tag{27}$$

$$= \sum_{i,j} \sum_{k=1}^{m} {c_{ij}}^2 \alpha_k^2 q_k^{ij} + \sum_{i,j} \sum_{1 \leq k < k' \leq m} 2 {c_{ij}}^2 \alpha_k \alpha_{k'} q_k^{ij} q_{k'}^{ij} \tag{28}$$

To drive Eq. 28 from Eq. 27, we can convert the square terms by using $\left(q_k^{ij}\right)^2 = q_k^{ij}$ because of $q_k^{ij}$ is 0 or 1. We can calculate the first term in Eq. 24 as the sum of linear and quadratic terms by substituting Eq. 28 into the first term of Eq. 26.

Now we can compare two sinograms $P(\theta, s)$ and $IP(\theta, s)$. To compute the energy minimization model, we subtract the values for each pixel in the two sinograms and square them.

$$F(\theta, s) = \sum_{\theta=0}^{180-d\theta} \sum_{s=1}^{n} \left( (IP - P)(\theta, s) \right)^2 \tag{29}$$

where $\theta$ is the projection angle, $s$ is the position of the sensor, and $d\theta$ is the amount of change in the projection angle. Now, $F(\theta, s)$ is expressed in linear terms, quadratic terms excluding constant terms. In the QUBO model, constant terms are excluded. The minimum value of the QUBO model is the opposite sign of the summation of constant terms $\sum_{\theta=0}^{180-d\theta} \sum_{s=1}^{n} \{P(\theta, s)\}^2$.

### 3.3 Preprocessing of X-ray projection images

A pixel value $I_{ij}$ in Eq. 20 can take from 0 to the maximum of X-ray mass attenuation coefficient of the sample. Therefore, the sinogram of the superposed state obtained by Radon transform has to starts from 0. In order to make the projection image created by Radon transformation include the values of the actual X-ray projection image, a constant value must be added to $I_{ij}$ or a constant value must be subtracted from the X-ray projection image. To express the X-ray coherence effect, more qubits must be allocated to $I_{ij}$. In this experiment, we will take the average of the space in the X-ray projection image and subtract it from the image (See Fig. 2d). Also, we will ignore the X-ray coherence effect. The sample is a tooth model made of resin. We made the sample out of a single material to minimize the use of logical qubits. In this case, $I_{ij}$ can be represented as $\alpha q^{ij}$.

The projection data for the sinogram used in our experiment are acquired from 0 to 180 degrees with the number of projections equal to the side length of the cross-sectional image to be reconstructed. For instance, for a $50 \times 50$ cross-sectional image, there are 50 projection angles. However, due to the possibility of various noise levels occurring at different angles during object scanning, the projection images may not be smoothly connected across each angle. Therefore, a preprocessing process is necessary to mitigate the discontinuities in the sinogram and ensure smoother connections. The first step of the preprocessing process is to normalize the pixel values of the sinogram according to the values of the first projection image. Next, because the X-ray mass attenuation coefficient $\alpha$ of a single material is constant, we can scale the entire sinogram to make $\alpha$ equal to 1, simplifying our QUBO formulation. The scaled sinogram can be obtained by applying a classical reconstruction algorithm to the sinogram from the first step to reconstruct the image, and then multiplying the sinogram by the average pixel value of the material region, calculated by dividing the sum of the pixel values by the number of pixels in that region. In our experiments with actual data samples, we created and used the scaled sinogram through this two-step preprocessing process.

### 3.4 Limitation of classical algorithm in the presence of data errors and/or limited information

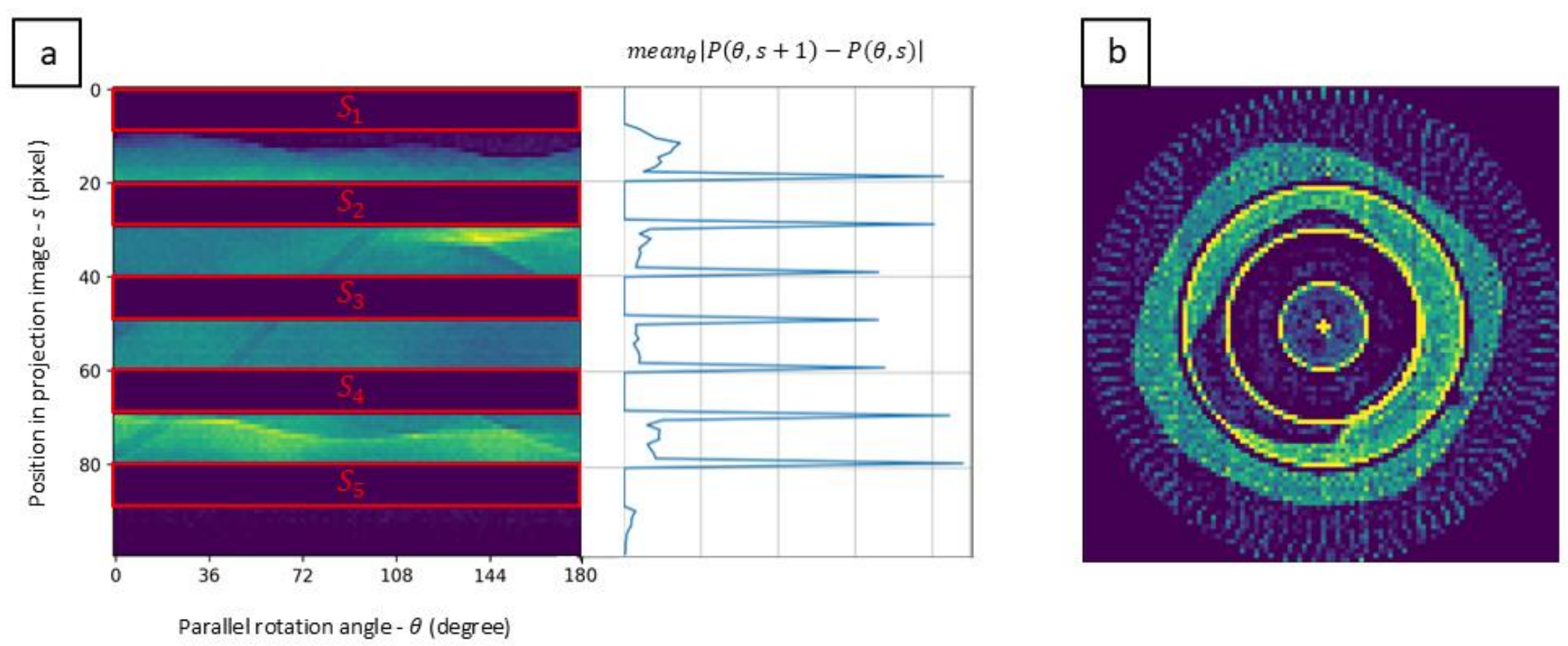

Figure 3. Illustration of the artificial horizontal error areas in the sinogram that can cause ring artifacts in the reconstructed image, along with the reconstruction results from this sinogram. (a) A sinogram with artificial errors and the graph of the average differences for each consecutive row. (b) A reconstructed image from the sinogram by the FFT.

To demonstrate the quantum supremacy of the quantum optimization tomographic reconstruction algorithm, two types of experiments were conducted. The first experiment involves removing ring artifacts caused by noise during the CT scanning in classical algorithms. To clearly demonstrate the difference between quantum and classical algorithms, we introduce artificial errors into the sinogram, as shown in Fig. 3a. When such errors consistently occur across different angles, circular patterns appear in the reconstructed image, as shown in Fig. 3b. However, our algorithm focuses on each pixel of the image to be reconstructed when formulating the QUBO, allowing us to address the issue by excluding the unreliable areas with errors in the sinogram during the QUBO formulation. Therefore, to apply our algorithm to a sinogram containing these horizontal error areas, it is necessary to first identify the positions in the projection images where the horizontal error areas occur. This can be achieved by identifying positions in the projection image where the density changes abruptly. Specifically, by creating a density difference graph using the differences between adjacent positions in the density graph of a single projection image and then averaging this across all θ, we obtain the graph represented by the blue line in Fig. 3a. By examining this graph, we can observe that the density difference changes abruptly at the beginning and end of the horizontal error areas. We define the union of the positions of these identified horizontal error areas as S and exclude the pixel values in these areas from the sinogram when generating the QUBO. Consequently, our cost function $F$ is defined as follows:

$$F(\theta,s) = \sum_{\theta=0}^{180-d\theta} \sum_{s\in\{1,\cdots,n\}\backslash S} \left((IP-P)(\theta,s)\right)^2$$

The second experiment to demonstrate quantum supremacy for our algorithm involves using projection images captured only within a smaller angular range, rather than the full 0° to 180°. If the maximum projection angle is denoted as $D°$ (where $D° < 180°$), then for the angle variable $\theta \in \{Md\theta | M \in \mathbb{Z}^{+} \cup \{0\}\ and\ Md\theta < D°\}$ the cost function $F$ in this case is given by the following:

$$F(\theta,s) = \sum_{\theta=0}^{Md\theta} \sum_{s=1}^{n} \left((IP-P)(\theta,s)\right)^2$$

The results of the two experiments conducted to demonstrate quantum supremacy can be found in Sections 4.2 and 4.3.

## 4. RESULTS

### 4.1 Segmentation

In this paper, we compare the segmented images obtained from the classical CT image and the segmentation image generated using a quantum annealer. For the classical segmented image, we generated a sinogram corresponding to the axial level indicated by the red line in Fig. 4a (See Fig. 4b). The CT image shown in Fig. 4c was reconstructed using the fast Fourier transform algorithm provided by MATLAB. Subsequently, we applied a MATLAB binarization algorithm to the CT image to identify the interior of the tooth, as depicted in Fig 3d. Finally, we obtained a segmented image of axial level 16 for the tooth sample (See Fig. 4e).

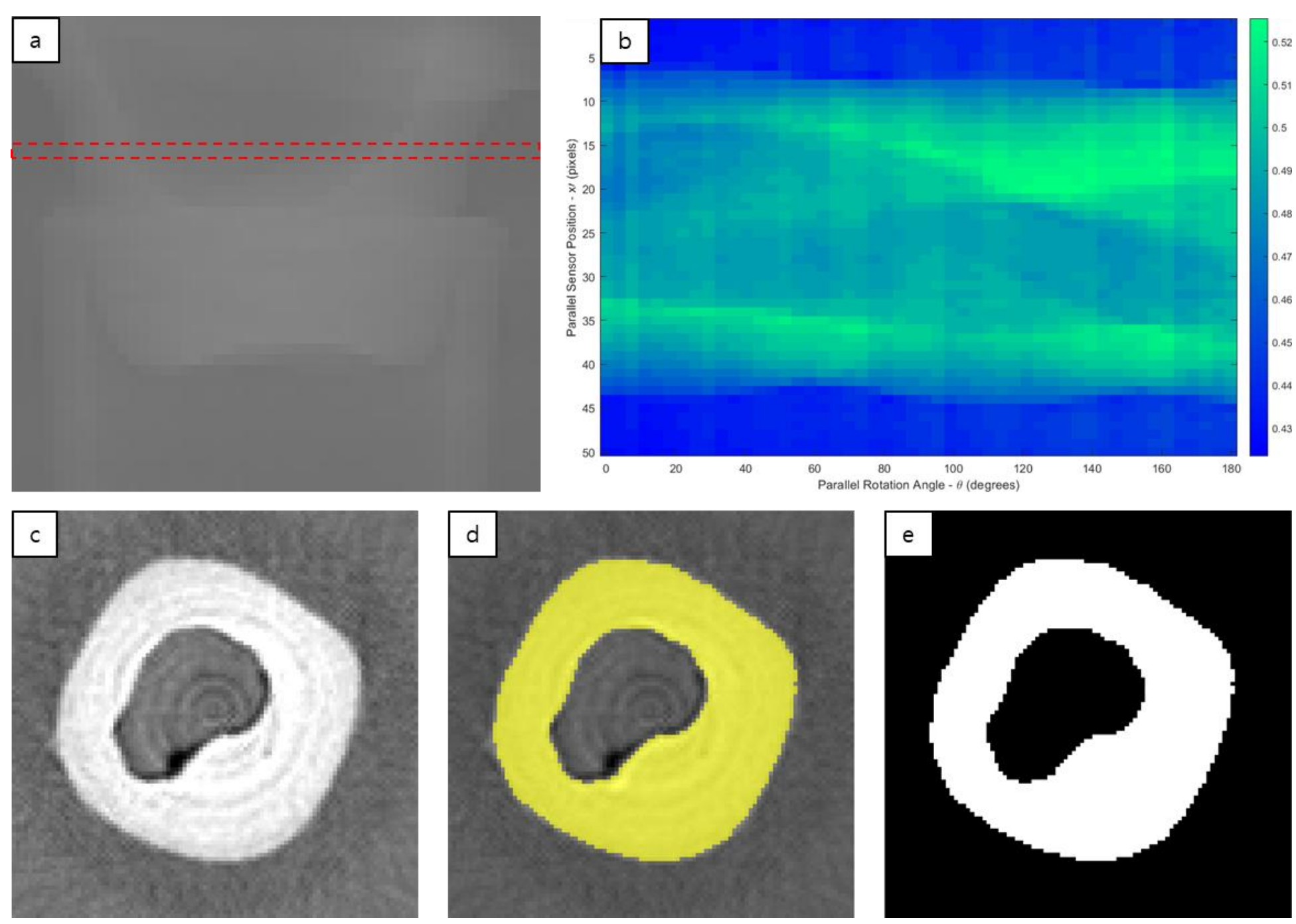

Figure 4. Classical CT image segmentation. (a) The axial level used for CT image reconstruction is displayed in red on the X-ray projection image. (b) This figure is a sinogram obtained at axial level 16. (c-e) A CT image and segmented image reconstructed by classical algorithm.

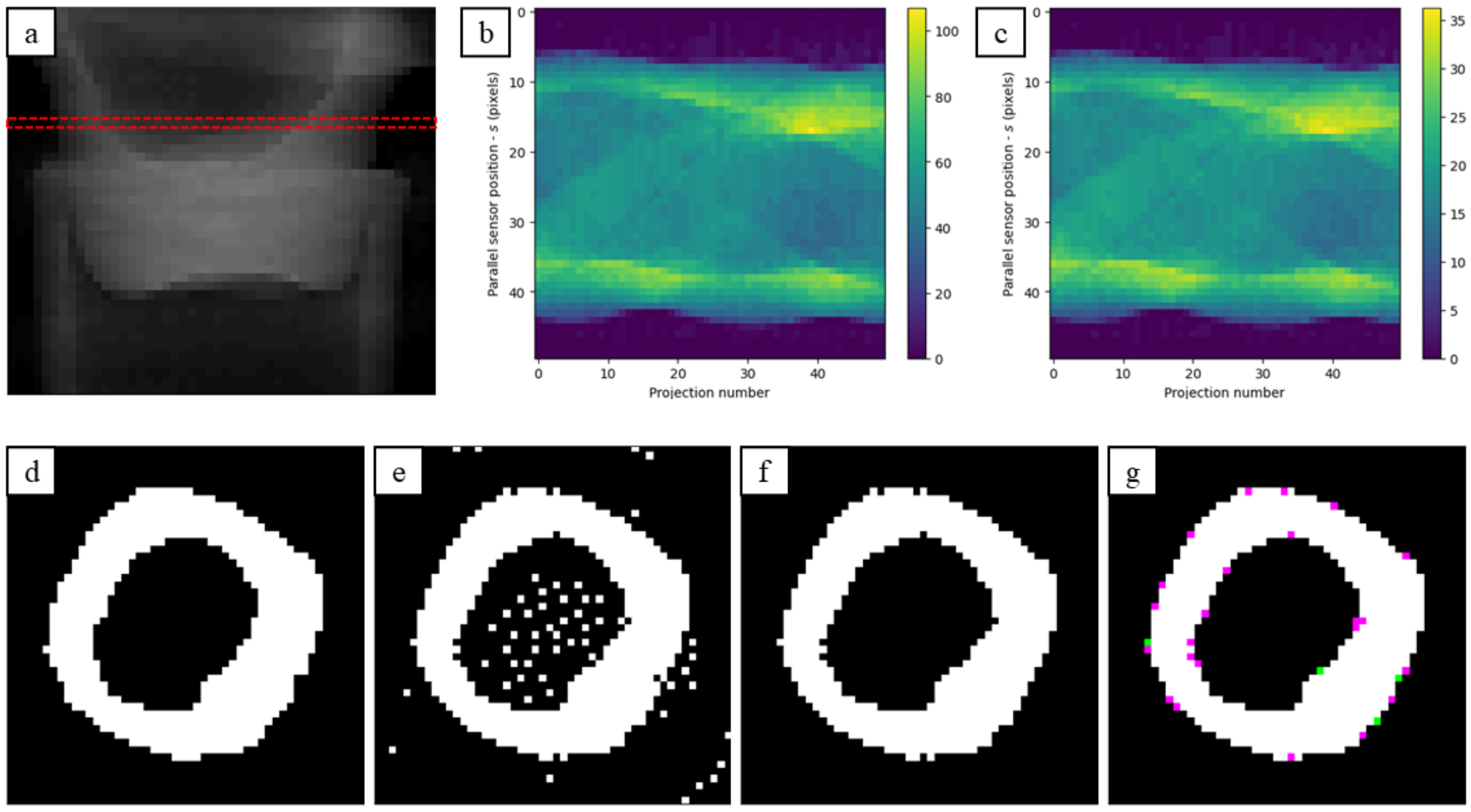

Figure 5. A 50 by 50 CT image segmentation by quantum optimization algorithm. (a) The axial level used for CT image reconstruction is displayed in red on the X-ray projection image. (b) This figure is a sinogram obtained at axial level 14. (c) A scaled sinogram obtained by adjusting the entire sinogram to make the X-ray mass attenuation coefficients close to 1, after scaling all projection images according to the range of the first projection image. (d) The image segmented an image reconstructed from the original sinogram using the classical reconstruction method by the classical image binarization technique. (e) A segmented image by our quantum optimization algorithm from the scaled sinogram c. (f) A postprocessed image from our segmented image e. (g) This image compares two segmented images obtained by classical and quantum algorithms.

We utilize the hybrid solver of the D-Wave system to execute the quantum optimization tomographic reconstruction algorithm. This novel quantum algorithm enables the computation of segmented images directly from X-ray data in a single step. Here, we conducted experiments using a sinogram obtained by binning into 50 by 50. As illustrated in Fig. 5b, a sinogram is derived from the projection image, which is proportionate to the X-ray mass attenuation coefficient displayed in Fig. 5a. However, varying noises are present with different projection angles, resulting in sporadic dark vertical lines in the sinogram, as shown in Fig. 5b, rendering the sinogram overall not smooth. We applied the preprocessing process to the original sinogram to smooth it and created a scaled sinogram, allowing us to assume the X-ray mass attenuation coefficient as 1 (as seen in Fig. 5c). To validate the results of our quantum segmentation algorithm, we utilize the sinogram from Fig. 5c to generate a CT image in Fig. 5d using a classic algorithm and use this CT image as the solution CT image. This image is segmented using the classical image binarization technique using a threshold after reconstructing the image from the original sinogram Fig. 5b using the fast Fourier transform. Here, the threshold value was set as the maximum value among the pixel values of some areas of the empty space inside the teeth in the reconstructed image, and the areas were made to disappear to 0 through image binarization. Now, let's apply our quantum segmentation algorithm to the rescaled sinogram in Fig. 5c. By formulating the QUBO model and calculating the pixel values of the sinogram Eq. 27, we aim to represent all values through the superposed CT image. The theoretically calculated global minimum energy is determined as the negative sum of the squared values of each pixel in the sinogram, which, in our dataset, amounts to $-756233.5433480422$. Employing the hybrid solver for the QUBO model, we obtain a minimum energy value $-753748.191693$. Figure 5e showcases the segmented image corresponding to the minimum energy value. Then, to create a cleaner CT image, we employed post-processing methods utilizing pixel connectivity to fill black holes and erase white dots, resulting in Fig. 5f. Furthermore, Fig. 5g presents a comparative analysis between the segmented image derived from the classical algorithm in Fig. 5d and the segmented image obtained through the quantum optimization algorithm in Fig. 5f. If you look closely, you can see that errors occur only at the border between the teeth and the background.

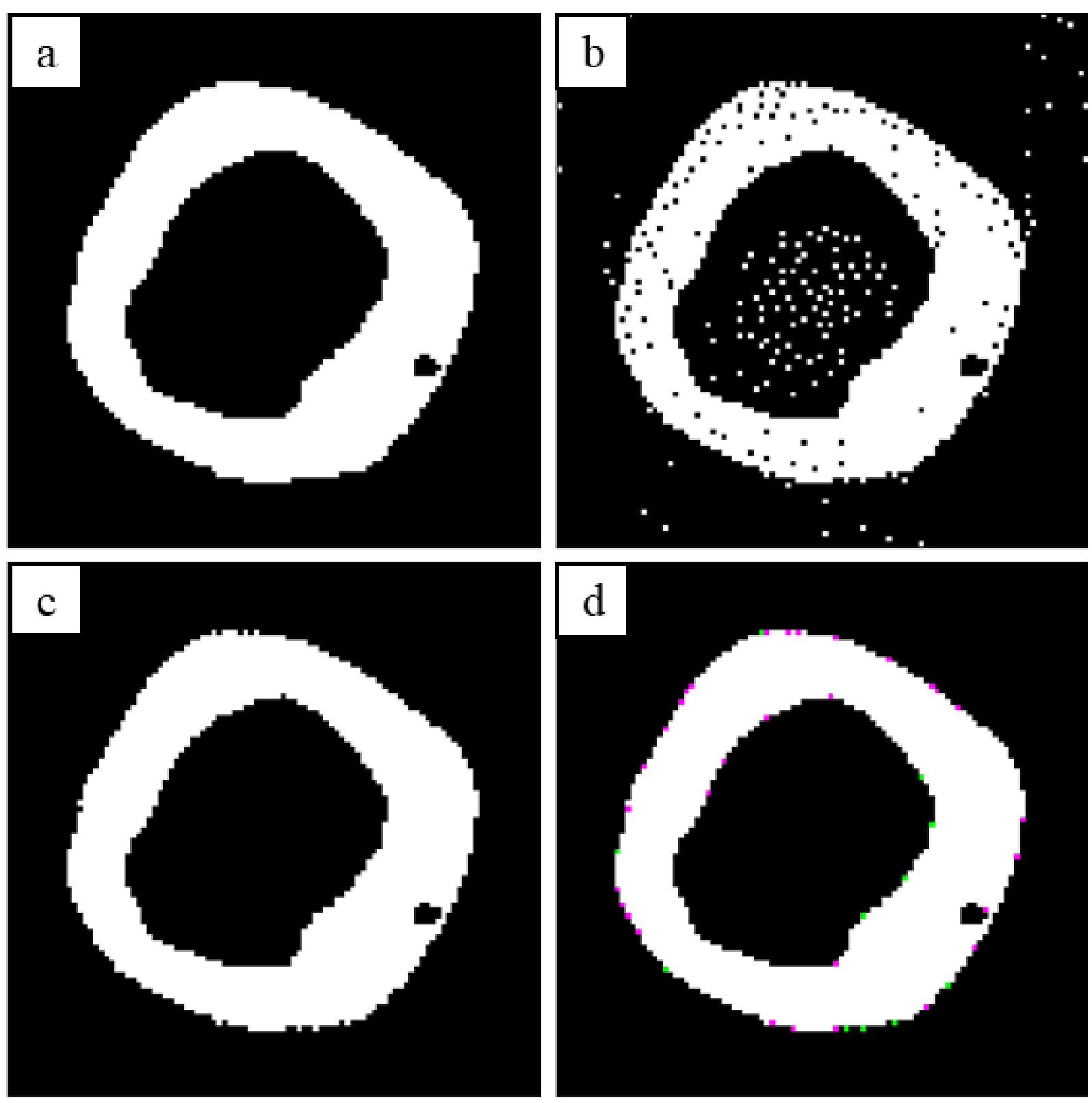

Figure 6. A 100 by 100 CT image segmentation by quantum optimization algorithm. (a) A reconstruction image from the original sinogram by the classical segmentation algorithm. (b) A segmented image by our quantum optimization algorithm from the rescaled sinogram. (c) A postprocessed image from b. (d) This image compares two segmented images obtained by classical and quantum algorithms.

We also conducted experiments using a sinogram obtained by binning into 100 by 100, which has a higher resolution than 50 by 50. Following the same procedure as with the 50 by 50 binning, we presented images segmented using FFT and classic algorithms on the original sinogram in Fig. 6a. The theoretically calculated global minimum energy for this image is $-9777411.550436938$. Using the rescaled sinogram and our hybrid solver, we obtained a minimum energy value of $-9730916.296585$. The segmented image obtained at this point is shown in Fig. 6b. In comparison, the image with removed dots and holes is presented in Fig. 6c. Additionally, we compared the segmentation results using classic and quantum algorithms, displayed in Fig. 6d. Similar to the case with 50 by 50 binning, we observe differences only at the boundaries between teeth and background even with 100 by 100 binning. However, it is evident that with the increase in resolution, more accurate segmentation becomes achievable.

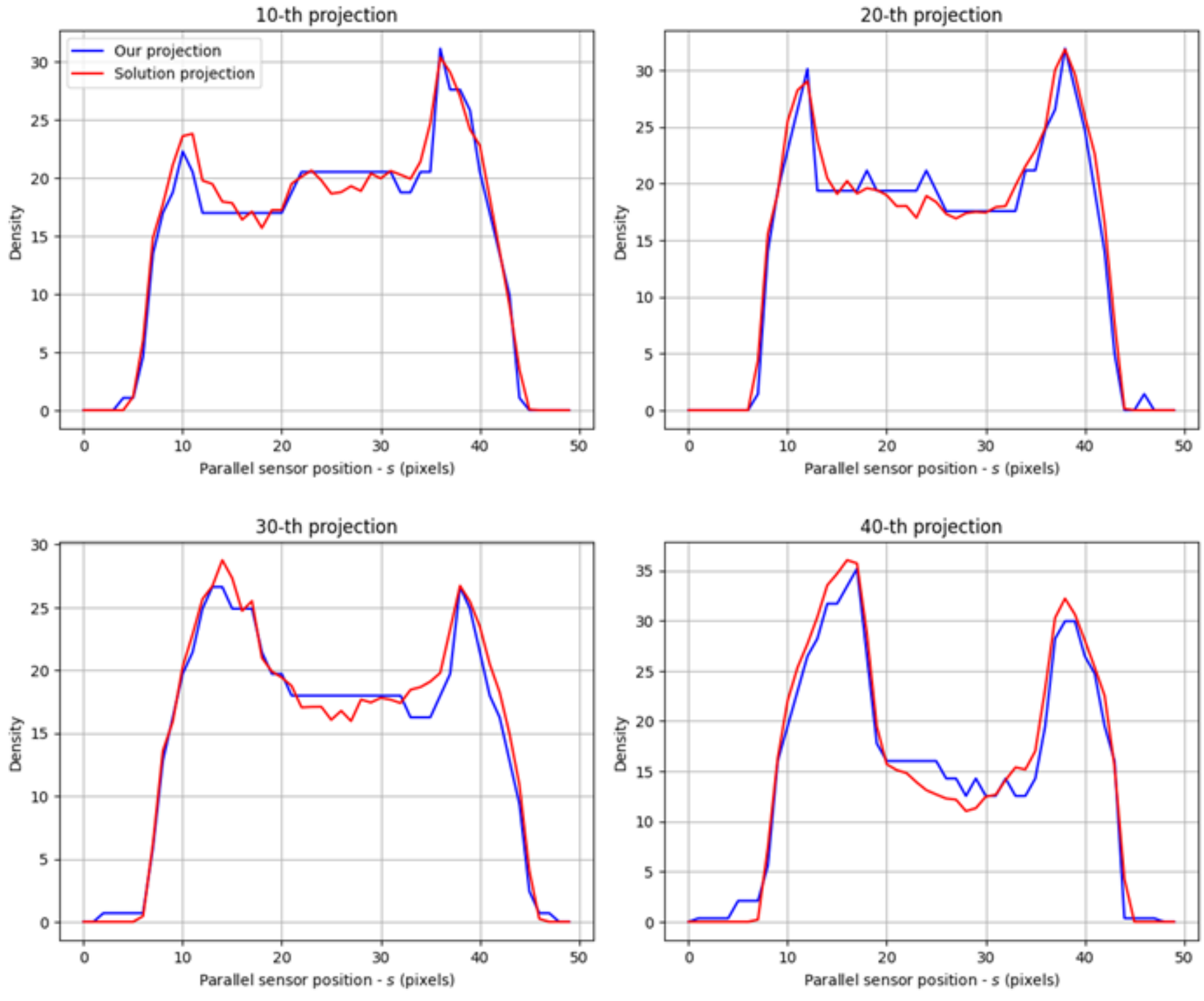

Figure 7. Overlapped density graphs of two projection images at 10, 20, 30, and 40th projection angles. Red line (labeled "Solution projection") depicts the projection image's density of a sinogram obtained by the solution CT image in Fig. 5d. Blue line (labeled "Our projection") depicts the projection image's density of the scaled sinogram in Fig. 5c.

To understand why black holes and white dots appear in the segmentation results obtained using the quantum segmentation algorithm, we compared the sinogram generated from the segmentation image (Fig. 5d) with the rescaled sinogram (Fig. 5c) used for our quantum segmentation algorithm. Figure 7 shows a graph superimposing the density of the 10th, 20th, 30th, and 40th projection images among the 50 projection images of the two sinograms mentioned above. The red line represents the density of the projection images made from the solution CT image where the segmented image from the CT image with FFT algorithm, while the blue line represents the density of the rescaled projection images we used. The general trends of the two graphs are similar, but differences between the graphs can be observed between the two peaks. Additionally, there are parts at the ends of the red graph corresponding to the background outside the teeth in the CT image, which is zero, whereas in the blue graph, there are non-zero values in the same area. The various errors in actual X-ray images make it difficult to achieve consistent densities across projection images, as they may perceive the almost same X-ray mass attenuation coefficients differently at different pixel locations within the teeth. Therefore, simply rescaling the pixel values of the projection images to create grayscale images with consistent interiors is challenging. It is considered that these small errors in the projection images contribute to the generation of dots and holes in the results of the quantum segmentation algorithm.

### 4.2 Quantum supremacy results using projection data with noise

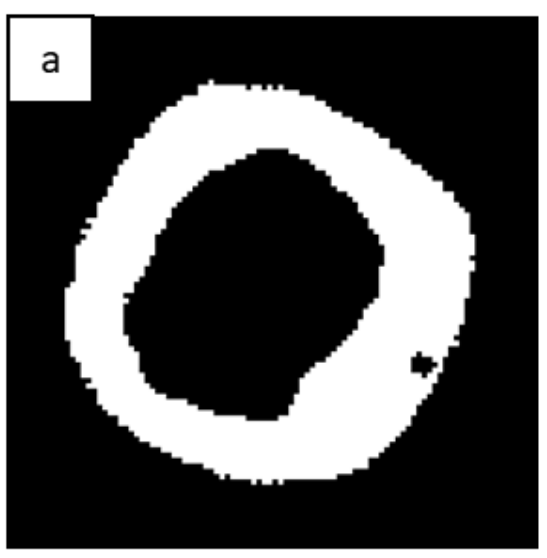

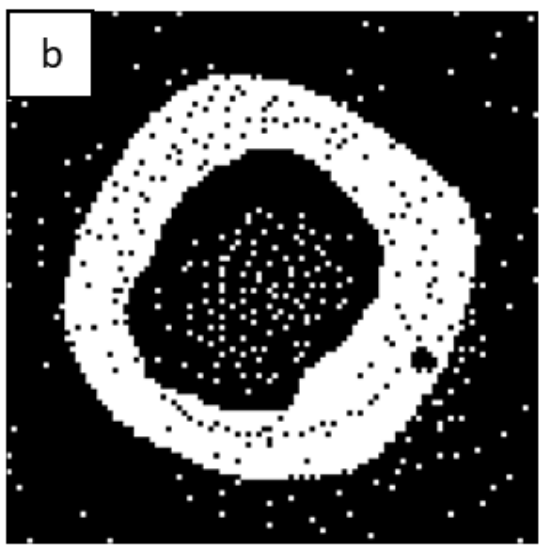

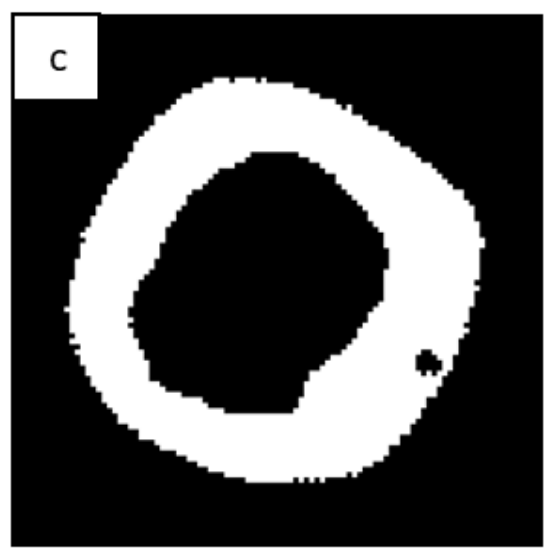

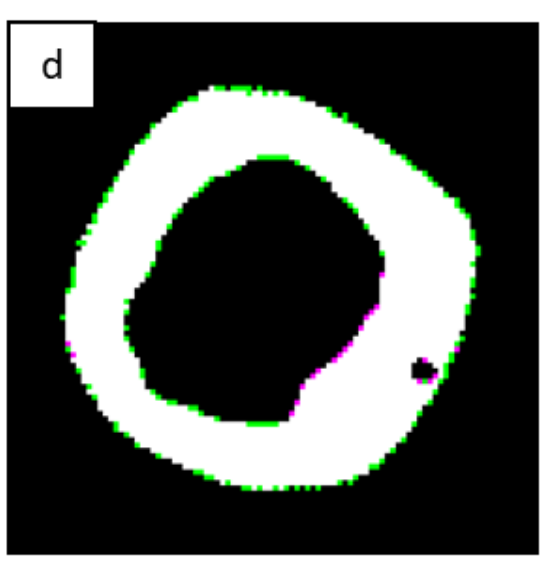

Figure 8. Results of applying our algorithm to the sinogram with five artificially inserted horizontal error areas, as shown in Fig. 3a, for reconstructing a $100 \times 100$ cross-sectional image of the tooth sample. (a) Solution image obtained by applying FFT and thresholding techniques to the original sinogram before adding artificial ring artifacts. (b) A segmented image obtained by our algorithm and hybrid solver from the sinogram in Fig. 3a. (c) A postprocessed image from (b). (d) The result of overlapping the solution image with our result in (c).

We artificially introduce ring artifacts into the sample to demonstrate quantum supremacy of the quantum optimization tomographic algorithm. The experiments were conducted under controlled conditions to ensure that the impact of these artifacts on the algorithm's performance could be accurately assessed. Figure 3a illustrates the sinogram used in the experiment to artificially generate ring artifacts, along with the corresponding results. The images used in the experiment represent a $100 \times 100$ cross-section of a tooth sample at the 27th axial level, with projection angles uniformly divided into 100 intervals from 0° to 180°. To create artificial artifacts, zero-rows with a length of 10 and a spacing of 10 were introduced into the sinogram, as shown in Fig. 3a. The result of applying FFT to this sinogram is displayed in Fig. 3b. To make the ring artifacts more visible, the resulting image was constrained to pixel values between 0 and 1. When our algorithm was applied to the same sinogram, it yielded results similar to those shown in Fig. 6b, where the major structures are well reconstructed but with numerous visible black and white dots. After undergoing postprocessing, the resulting image, shown in Fig. 8c, displays a well-segmented cross-section of the tooth. In fact, as seen in Fig. 8d, when overlapping the postprocessed image with the solution image, differences can be observed only at the boundaries of the material.

### 4.3 Quantum supremacy results for limited projection angles

To demonstrate the superiority of our algorithm with smaller angular ranges, we set the maximum angle to be less than 180° in this experiment. The cross-sectional image of the tooth sample was set to $50 \times 50$ pixels, and the axial level was 15. To address concerns about using too few projection angles due to the smaller angular range, we initially divided the range from 0° to 180° into 100 angles and created 100 projection images. To create the solution image, FFT and thresholding methods were applied to the entire sinogram (see Fig. 9d). First, we conducted an experiment using initial 50 out of the 100 projection images, corresponding to angles smaller than 90°. However, Fig. 9e shows the result of applying our algorithm, the top-left portion exhibits significant errors. To explore the impact of using more angles, we conducted a second experiment using 60 projection images, which corresponds to angles smaller than 108°. The results of applying FFT to the 50 and 60 projection images are shown in Fig. 9b and Fig. 9c, respectively. In both cases, the FFT results show that the pixel values deviate from the range of 0 to 1, and the boundary shapes are indiscernible. In contrast, as seen in Fig. 9e and Fig. 9h, our method preserves the overall structure. Although errors are detected in the top-left and bottom-right areas, as shown in Fig. 9g and Fig. 9j when compared with the solution image after post-processing, it is evident that using 60 projection images results in fewer discrepancies with the solution image compared to using 50 projection images.

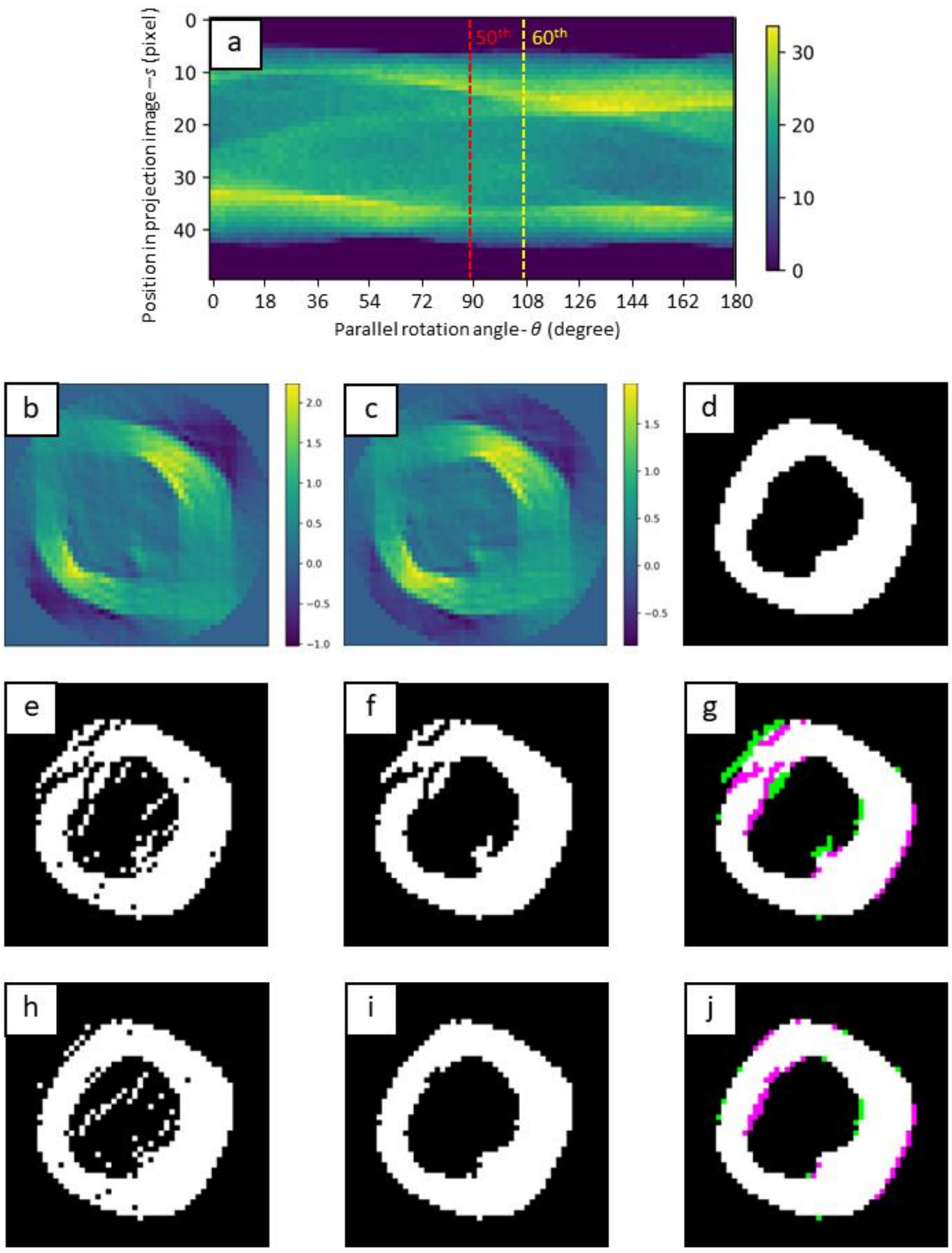

Figure 9. Experimental results on the limited angles for a $50 \times 50$ cross-sectional image at axial level 15 of the tooth sample. (a) Sinogram created by evenly dividing 0 to 180 degrees into 100 parts. (b) Result of applying FFT using only the first 50 projection images in the sinogram (a). (c) Result of applying FFT using 60 projection images. (d) Solution image obtained by applying FFT to the entire sinogram and using a thresholding method. (e-g) Results obtained by applying our algorithm to the 50 projection images using a hybrid solver, post-processing results, and the result of overlapping with the solution image. (h-j) Results obtained by applying our algorithm to the 60 projection images using a hybrid solver, post-processing results, and the result of overlapping with the solution image.

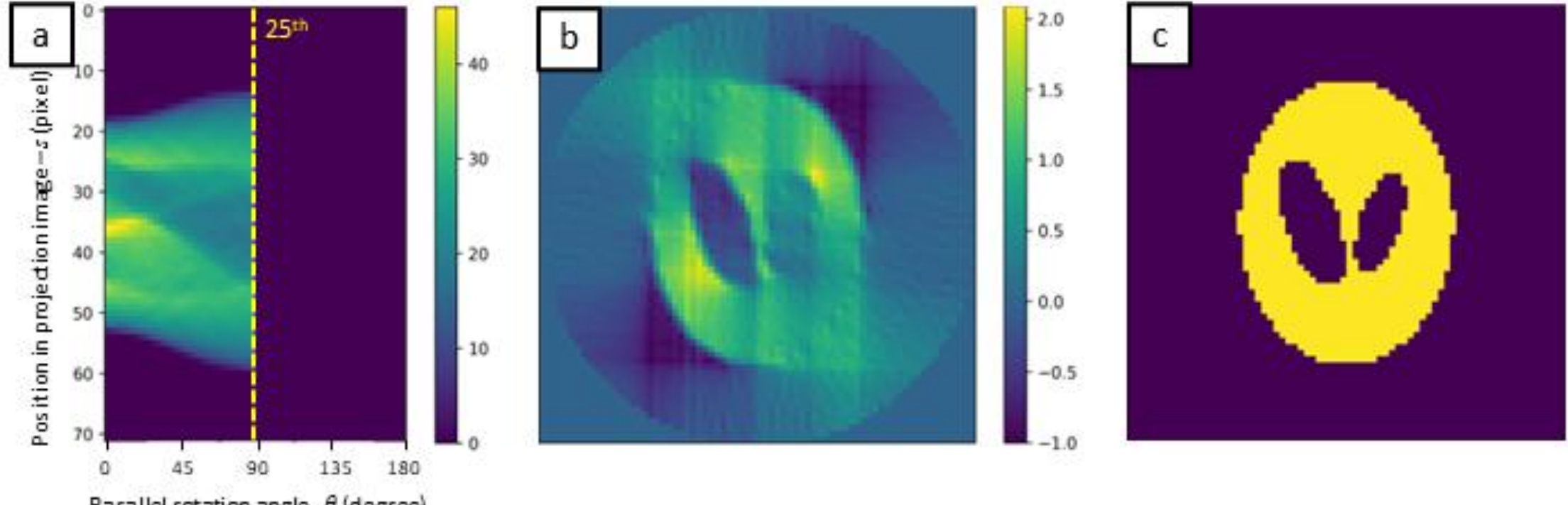

Figure 10. Experimental results on the limited angles for a 50 × 50 Shepp-Logan binary sample image with zero-pads. (a) Sinogram created using only the first 25 projection images (under 90°) out of 50, which were evenly divided from 0° to 180°. (b) Result of applying FFT to the sinogram in (a). (c) Result of applying our algorithm to the sinogram in (a). This figure is also the original sample image used in the experiment.

To conduct an experiment on a noise-free sinogram, we also tested using a 50x50 Shepp-Logan binary sample image with a zero pad of thickness 11 added on all sides (see Fig. 10c). In this case, as shown in Fig. 10a, we limited the projection angles to 90 degree and used only 25 projection images, which is half of the total 50. The result of applying FFT to these 25 projection images, as seen in Fig. 10b, shows that the outer boundaries in the top-left and bottom-right corners, as well as the right ellipse among the two inner ellipses, are difficult to discern. The result of applying our algorithm to this case is shown in Fig. 10c, and it exactly matches the sample image used in the experiment.

## 5. DISCUSSION

The quantum algorithm for CT image reconstruction[15] necessitates a number of qubits equal to the number of bits in each pixel. To reconstruct a 50 by 50 CT image with a resolution of 64 bits, a total of 160,000 logical qubits are required. In contrast, the quantum optimization tomographic reconstruction algorithm can reconstruct and segment the CT image with only 2,500 logical qubits, provided that the X-ray mass attenuation coefficients of the sample are known. This new algorithm enables the generation of more precise segmented CT images while using fewer logical qubits. Figure 4b represents a sinogram derived using MATLAB, which was subsequently converted to an integer value in Python. Python's sinogram values and the Radon transform applied to the segmented image in Fig. 4e were utilized to calculate the X-ray mass attenuation coefficient of the sample. For this experiment, an approximate coefficient was employed, but we anticipate achieving more accurate results by acquiring the coefficient through experimental means. Due to the utilization of the X-ray mass attenuation coefficient in the new algorithm, attaining a theoretically achievable global minimum energy is not feasible. This is because the pixels encompassing the sample's shell contain a mixture of both the sample and the surrounding space, making it impractical to represent them solely based on the X-ray mass attenuation coefficient of each pixel. However, during the segmentation process, these pixels are transformed into binary values of 0 or 1, and the new algorithm effectively addresses this issue. The sinogram is crucial for the optimal use of the quantum optimization tomographic reconstruction algorithm. Accordingly, sinograms obtained from real samples must be processed to ensure smooth connectivity across projection angles. Therefore, meticulous image processing of real samples is necessary to achieve the best results with our quantum optimization tomographic reconstruction algorithm. However, currently, sinograms generated using our current rescaling method may yield noisy results. Nevertheless, through image post-processing, sufficiently clean segmentation results can be obtained. Comparing Figs. 4 and 5 using this method, it becomes clear that as image resolution increases, our algorithm produces more accurate segmentation results. Due to the limited number of qubits available in quantum computers, we can only process images of allowable sizes within that constraint. However, considering the rapid advancement of quantum annealers, our algorithm will enable image segmentation for high-resolution images in the coming years. We are also currently advancing this experiment to perform studies involving samples with multiple X-ray mass attenuation coefficients.

In general, digitized images have an error within half a pixel, and in order to have an error of less than 50μm, the size of each pixel is approximately 100μm. Human teeth exist within a field of view of approximately 20cm, and a single slice CT image of 2000 by 2000 size is required to display them with an accuracy of 50μm or less. Even if the quantum superposition phenomenon of the two X-ray mass attenuation coefficients for hard and soft tissues is used for each pixel, approximately $8 \times 10^6$ logical qubits are required to reconstruct the CT image for one axial level. If motion artifacts of the sample occur during scanning in a cone beam CT (CBCT) system, CT image reconstruction of the entire three-dimensional image is required, which requires approximately $16 \times 10^9$ logical qubits. The D-Wave system's hybrid solver provides up to 1 million binary variables that serve as logical qubits. Hybrid solvers also have the disadvantage of not having enough binary variables and having difficulty finding the global minimum energy when using many binary variables. Additionally, when calculating the QUBO model using Python, there are many limitations because the coefficients for linear and quadratic terms are used as dictionary types. If the hybrid solver calculates the QUBO model in the form of a QUBO matrix, it will be able to be freer from memory shortages when calculating large problems. Nevertheless, the quantum optimization tomographic reconstruction algorithm has the potential to find global minimum energy and fast calculations, and quantum annealers are also developing at a rapid pace, so they have tremendous potential.

When a sinogram containing horizontal error areas is reconstructed using a classical algorithm, ring artifacts are generated in the reconstructed image. However, our algorithm avoids this issue by focusing on each sinogram's pixel that is less affected by the error, and generating the QUBO model. This approach allows us to exclude unreliable regions from the sinogram while still incorporating qubit information from other reliable areas. Therefore, even when excluding unreliable areas from the sinogram to generate the QUBO, qubit information can still be obtained from the other reliable areas. However, when applying our algorithm after removing the unreliable regions, as shown in Fig. 8b, there are more black and white dots compared to the result without removing, as shown in Fig. 6b. This increase is likely because the original data obtained from scanning the actual tooth sample was not ideal, so the quantum algorithm failed to find the global minimum value. Nonetheless, our algorithm achieved better results when there were fewer areas with errors. Given these advancements, our algorithm's potential extends beyond CT scans. For instance, since electron tomography is typically performed within an angular range from $-70°$ to $70°$, our algorithm's ability to operate effectively within a $140°$ angular range represents a significant advantage[18]. Figure 9 shows that even with the $50 \times 50$ cross-sectional image for an actual tooth sample, we were able to successfully obtain a meaningful segmentation result that allows us to predict the shape using 60 angles below $108°$. Moreover, the experimental results for the $50 \times 50$ Shepp-Logan binary sample image produced a perfect segmentation result with just 25 angles below $90°$. While traditional classical algorithms often struggle with reconstruction in angular ranges smaller than $180°$, our algorithm performs effectively even within these reduced angular ranges, making it highly applicable to electron tomography.

This further underscores the algorithm's robustness and precision, enhancing its potential for application in electron tomography, where accurate reconstruction is crucial. Furthermore, we anticipate significant advancements in medical imaging diagnostics, particularly in areas like breast cancer and cardiovascular diagnosis, by substituting classical optimization algorithms with quantum optimization algorithms to address the low clarity of cone beam CT images.

## DATA AVAILABILITY

The Python code used in this paper is in the supplementary material.

## AUTHOR CONTRIBUTIONS STATEMENT

K. Jun conceived and designed the research problem. K. Jun performed the experiments. All authors developed theoretical results and wrote the paper.

## COMPETING FINANCIAL INTERESTS

The authors declare no competing financial or nonfinancial interests.

## ACKNOWLEDGE

PBCT was conducted at beamline 6C of the Pohang Light Source-II, which is funded by the Ministry of Science and ICT (MSIT) and POSTECH in Korea. This research was supported by the quantum computing technology development program of the National Research Foundation of Korea (NRF) funded by the Korean government (Ministry of Science and ICT (MSIT)) (No. 2020M3H3A111036513). Special thanks to Hyunju Lee's assistant.

[13] Schomberg, H. and Timmer, J. "The gridding method for image reconstruction by Fourier transformation", IEEE transactions on medical imaging 14.3 (1995) 596-607.

[14] Singh, R., Wu, W., Wang, G. and Kalra, M.K. "Artificial intelligence in image reconstruction: the change is here", Physica Medica 79 (2020) 113-125.

[15] Jun, K. "A highly accurate quantum optimization algorithm for CT image reconstruction based on sinogram patterns", *Scientific Reports*, *13*(1), 14407, (2023).

[16] Kim, B.C., Lee, H. and Jun, K. "Noise-resistant reconstruction algorithm based on the sinogram pattern", arXiv preprint arXiv:2111.10067 (2021).

[17] Lee, H., Ryu, Y., Kim, J. and Jun, K. "Quantum computing for the optimization of CT image reconstruction." 2022 13th International Conference on Information and Communication Technology Convergence (ICTC). IEEE, 2022.

[18] Ercius, P., Alaidi, O., Rames, M. J., & Ren, G., "Electron tomography: a three-dimensional analytic tool for hard and soft materials research", *Advanced materials*, *27*(38), 5638-5663 (2015).

[19] Jun, K. and Yoon, S. "Alignment Solution for CT Image Reconstruction using Fixed Point and Virtual Rotation Axis", Sci. Rep. 7, 41218 (2017).

[20] Jun, K. and Kim, D. "Alignment theory of parallel-beam computed tomography image reconstruction for elastic-type objects using virtual focusing method", Plos One, 13(6), e0198259 (2018).

[21] Jun, K. and Jung, J. "Virtual multi-alignment theory of parallel-beam CT image reconstruction for elastic objects", Sci. Rep. 9 (2019) 6847.

[22] Jun, K. "Virtual multi-alignment theory of parallel-beam CT image reconstruction for rigid objects", Sci. Rep. 9 (2019) 13518.

[23] Lim, J.H., et al. "Station for structural studies on macro objects: beamline 6C Bio Medical Imaging the Pohang Light Source-II", Biodesign. Vol.5, No.2, 53-61 (2017).

[24] Hubbell, J. H. "Photon mass attenuation and energy-absorption coefficients", Int. J. Appl. Radiat. Isot. 33, 1269–1290 (1982).